\documentclass[12pt,showpacs,showkeys,amsmath,amssymb]{revtex4}
\usepackage{amsmath,amsfonts,amsthm,amscd,amssymb,latexsym}
\usepackage{bm}
\usepackage{dcolumn}
\usepackage{graphicx}
\usepackage{epstopdf}
\usepackage{color}
\usepackage{epsf}
\usepackage{epsfig}
\usepackage{graphicx, epic, eepic, color}

\def\@{\partial_}

\def\negenspace{\kern-1.1em}

\def\sqr#1#2{{\vcenter{\hrule height.#2pt\hbox{\vrule width.#2pt
height#1pt \kern#1pt \vrule width.#2pt}\hrule height.#2pt}}}

\date{\today}

\begin{document}
\title{Nonlocal Gravity in the Solar System}

\author{C. Chicone}
\email{chiconec@missouri.edu}
\affiliation{Department of Mathematics and Department of Physics and Astronomy, University of Missouri, Columbia,
Missouri 65211, USA }

\author{B. Mashhoon}
\email{mashhoonb@missouri.edu}
\affiliation{Department of Physics and Astronomy,
University of Missouri, Columbia, Missouri 65211, USA}

\begin{abstract} 

The implications of the recent classical nonlocal generalization of Einstein's theory of gravitation for gravitational physics in the Solar System are investigated. In this theory, the nonlocal character of gravity appears to simulate dark matter. Nonlocal gravity in the Newtonian regime involves a reciprocal kernel with three spatial parameters, of which two have already been determined from the rotation curves of spiral galaxies and the internal dynamics of clusters of galaxies. However, the short-range parameter $a_0$ remains to be determined.  In this connection, the nonlocal contribution to the perihelion precession of a planetary orbit is estimated and a preliminary lower limit on $a_0$ is determined. 

\end{abstract}



\maketitle

\section{Introduction}

Relativity theory contains a basic postulate of \emph{locality}, since Lorentz invariance is extended in a \emph{pointwise} manner to the measurements of accelerated observers in Minkowski spacetime. This same assumption accounts for the \emph{local} nature of Einstein's principle of equivalence, which implies that an observer in a gravitational field is \emph{locally} inertial~\cite{Ei}. However, classical field measurements are intrinsically nonlocal, since they generally involve a spacetime average of the field along the past world line of the observer~\cite{B+R1, B+R2, BM1}. Indeed, the field is always local, but satisfies integro-differential field equations. On this basis a nonlocal special relativity theory has been developed in which the locality postulate for fields is extended for accelerated observers by the inclusion of  certain averages of the fields over their past world lines with kernels that contain the memory of the observers' past accelerations~\cite{BM2}. The deep connection between inertia and gravitation suggests that gravitation could be history dependent as well.  In a series of recent papers, a nonlocal generalization of Einstein's theory of gravitation has been developed in which nonlocality is due to the gravitational memory of past events~\cite{NL1, NL2, NL3, NL4, NL5, NL5a, NL6, NL7}. In this nonlocal theory of gravity, the gravitational field is still local, but satisfies integro-differential equations that go beyond the field equations of general relativity via a causal kernel that represents the gravitational memory of past events. At the present stage of the development of nonlocal gravity, the nonlocal kernel must be determined from observation. The main purpose of this paper is to discuss, within the Newtonian regime of nonlocal gravity, the significance of observational data in the Solar System for the determination of the nonlocal kernel. 

In the Newtonian regime of nonlocal gravity, the theory reduces to a nonlocal modification of Poisson's equation for the gravitational potential $\Phi$. That is, let
\begin{equation}\label{I1}
\nabla^2\Phi (\mathbf{x}) = S (\mathbf{x})\,;
\end{equation}
then, the density of matter is the source of the gravitational potential through
\begin{equation}\label{I2}
S (\mathbf{x}) + \int \chi (\mathbf{x}-\mathbf{y})\,S (\mathbf{y})\, d^3y  = 4\pi G\,\rho(\mathbf{x})\,.\end{equation}
Here, $\chi$ is the universal convolution kernel of the theory in the Newtonian regime and $\rho$ is the matter density.  In nonlocal gravity, the nonlocal aspect of the gravitational interaction involves a certain causal spacetime average of the gravitational field. The corresponding kernel of the linear response encodes the persistent spacetime memory of the field. In the Newtonian regime, where the speed of light formally approaches infinity $(c\to \infty)$, retardation effects are totally absent and $\chi$ naturally involves only spatial gravitational memory. 

Under certain favorable mathematical conditions that are discussed below, Eq.~\eqref{I2} may be written as 
\begin{equation}\label{I3}
\nabla^2\Phi (\mathbf{x}) = 4\pi G\,\left[\rho(\mathbf{x})+\int q(\mathbf{x}-\mathbf{y})\, \rho(\mathbf{y})\,d^3y\right]\,,
\end{equation}
where $q$ is the \emph{reciprocal} convolution kernel~\cite{Tr, WVL} and 
\begin{equation}\label{I4}
 \rho_D(\mathbf{x}) :=\int q(\mathbf{x}-\mathbf{y})\, \rho(\mathbf{y})\,d^3y\,
\end{equation}
has the interpretation of the density of the \emph{effective} dark matter. The persistent negative result of experiments that have searched for the particles of dark matter naturally leads to the possibility that what appears as dark matter in astrophysics and cosmology is in fact an aspect of the gravitational interaction. 
The nonlocal character of gravity, however, cannot yet replace dark matter on all physical scales.  Indeed, dark matter is currently indispensable for explaining: (i) gravitational dynamics of galaxies and clusters of galaxies~\cite{Zw, RF, RW, SR, Sei, HMK}, (ii) gravitational lensing observations in general and the Bullet Cluster~\cite{BC1, BC2} in particular and (iii) the formation of structure in cosmology and the large scale structure of the universe. We emphasize that nonlocal gravity theory is so far in the early stages of development and only some of its implications have been confronted with observation~\cite{NL6}. Moreover, a beginning has recently been made in the development of nonlocal Newtonian cosmology~\cite{CCBM}. 

It follows from combining Eqs.~\eqref{I2} and~\eqref{I3} that kernels $\chi$ and $q$ are reciprocal to each other; that is,  two relations can in general  be deduced that reduce to the following \emph{reciprocity relation} 
\begin{eqnarray}\label{I5}
\chi(\mathbf{x}-\mathbf{y})+q(\mathbf{x}-\mathbf{y})+\int \chi(\mathbf{x}-\mathbf{z})\,q(\mathbf{z}-\mathbf{y})~d^3z=0\,
\end{eqnarray}
for convolution kernels. Indeed, in the integrand of Eq.~\eqref{I5}, the change of variable $\mathbf{z}$ to $\mathbf{u}$, via $\mathbf{z}-\mathbf{y}=\mathbf{x}-\mathbf{u}$, leads to the result that Eq.~\eqref{I5} is completely symmetric with respect to the interchange of $\chi$ and $q$. 

\subsection{Fourier Transform Method}

The transition from Eq.~\eqref{I2} to Eq.~\eqref{I3} can be implemented in the space of functions that are absolutely integrable ($L^1$) as well as square integrable ($L^2$) over all space. This has been demonstrated in detail in Ref.~\cite{NL5}. Let $\hat{s} (\boldsymbol{\xi})$ be the Fourier integral transform of a function $s(\mathbf{x})$ that is both $L^1$ and $L^2$; then, 
\begin{equation}\label{II1}
\hat{s} (\boldsymbol{\xi}) =  \int s(\mathbf{x})\, e^{-i\,\boldsymbol{\xi} \cdot \mathbf{x}}\, d^3x\,, \qquad 
s(\mathbf{x})=\frac{1}{(2\pi)^3}\,\int \hat{s} (\boldsymbol{\xi})\,e^{i\,\boldsymbol{\xi} \cdot \mathbf{x}}\, d^3\xi\,.
\end{equation}
It follows from the convolution theorem for Fourier integral transforms and Eq.~\eqref{I2} that
\begin{equation}\label{II2}
\hat{S}(\boldsymbol{\xi}) \,[1+\hat {\chi} (\boldsymbol{\xi})]= 4 \pi\,G \,\hat{\rho}(\boldsymbol{\xi})\,. \end{equation}
Similarly, it follows from Eq.~\eqref{I3} that 
\begin{equation}\label{II3}
\hat{S}(\boldsymbol{\xi})= 4 \pi\,G \,\hat{\rho}(\boldsymbol{\xi})\,[1+\hat {q} (\boldsymbol{\xi})]\,. \end{equation}
Combining Eqs.~\eqref{II2} and~\eqref{II3}, we find 
\begin{equation}\label{II4}
(1+\hat{\chi})(1+\hat{q})=1\,,
\end{equation}
which  is  reciprocity relation~\eqref{I5} expressed in the Fourier domain. It follows that if $q(\mathbf{x})$ is given by experimental data regarding dark matter, see Eq.~\eqref{I4},  and subsequently $\hat{q}(\boldsymbol{\xi})$ is calculated from the Fourier integral transform of $q(\mathbf{x})$, then the kernel of nonlocal gravity  $\chi(\mathbf{x})$ can be determined from the Fourier transform of $\hat{\chi}(\boldsymbol{\xi})$ that is given by Eq.~\eqref{II4}, namely, 
\begin{equation}\label{II5}
\hat{\chi}(\boldsymbol{\xi}) =- \frac{\hat{q} (\boldsymbol{\xi})}{1+\hat{q} (\boldsymbol{\xi})}\,, 
\end{equation}
provided
\begin{equation}\label{II6}
1+\hat{q} (\boldsymbol{\xi})\ne 0\,.
\end{equation}
Thus an acceptable reciprocal kernel $q(\mathbf{x})$ should be a smooth function that is $L^1$, $L^2$ and satisfies requirement~\eqref{II6}. We now proceed to the determination of $q(\mathbf{x})$.  

\subsection{Kuhn Kernel $q_K$}

The nonlocal Poisson Eq.~\eqref{I3} is in a form that can be compared with observational data regarding, for instance,  the rotation curves of spiral galaxies. Imagine, for instance, the circular motion of stars (or gas clouds) in the disk of a spiral galaxy about the galactic bulge. According to the Newtonian laws of motion, such a star (or gas cloud) has a centripetal acceleration of $v_0^2/r$, where $v_0$ is its constant speed; moreover,  this centripetal acceleration must be equal to the Newtonian gravitational acceleration of the star. Observational data indicate that $v_0$ is nearly the same for all stars (and gas clouds) in the galactic disk, thus leading to the nearly \emph{flat} rotation curves of spiral galaxies. This means that the ``Newtonian"  force of gravity varies essentially as $1/r$ on galactic scales. Attributing this circumstance to an effective density of dark matter and assuming spherical symmetry, it follows from 
\begin{equation}\label{III1}
\nabla \cdot [D(r)\, \hat{\mathbf{r}}]=\frac{1}{r^2} \, \frac{d}{dr}[r^2\, D(r)]
\end{equation}
that for $D=1/r$, we get from Poisson's equation of Newtonian gravity that the corresponding effective density of dark matter $\rho_D$ must be $v_0^2/(4\pi G r^2)$. Using Eq.~\eqref{I4} with $\rho(\mathbf{x})=M\,\delta(\mathbf{x})$, where $M$ is the effective
 mass of the galactic core, we find for kernel $q$,
\begin{equation}\label{III2}
q_K(\mathbf{x}-\mathbf{y})=\frac{1}{4\pi\lambda}\,\frac{1}{|\mathbf{x}-\mathbf{y}|^2}\,,
\end{equation}
where $\lambda=GM/v_0^2$ should be a (universal)  constant length of the order of 1\,kpc. 

It is remarkable that a modified Poisson equation of the form~\eqref{I3} with kernel~\eqref{III2} was suggested by Kuhn about 30 years ago;  in fact, it is interesting to digress briefly here and mention the phenomenological  Tohline-Kuhn  modified-gravity approach to the problem of dark matter~\cite{T1, T2, K1, K2}.  According to this scheme, the ``flat" rotation curves of spiral galaxies lead to a (Tohline-Kuhn)  modification of the Newtonian inverse-square law of gravity, namely,  
\begin{equation}\label{III3}
F_{TK}(r)=\frac{Gm_1m_2}{r^2}+ \frac{Gm_1m_2}{\lambda\, r}\,,
\end{equation}
where the relative deviation from Newton's law due to the long-range (``galactic") contribution is given by $r/\lambda$. In 1983, Tohline showed that this modification leads to the stability of the galactic disk~\cite{T1}. The gravitational potential for a point mass $M$ corresponding to this modified force law can be written as~\cite{T1} 
\begin{equation}\label{III4}
\Phi_T(\mathbf{x})=-\frac{GM}{|\mathbf{x}|}
+\frac{GM}{\lambda}\ln\left(\frac{|\mathbf{x}|}{\lambda}\right)\,.
\end{equation}
This Tohline potential satisfies Eq.~\eqref{I3} with  Kuhn kernel~\eqref{III2} when $\rho(\mathbf{x})=M\,\delta(\mathbf{x})$. 

The work of Kuhn and his collaborators contained a significant generalization of Tohline's original suggestion~\cite{K1, K2}; the Tohline-Kuhn scheme has been admirably reviewed by Bekenstein~\cite{B}.

\subsection{Derivation of Reciprocal Kernel $q$}

The reciprocal kernel must satisfy certain mathematical requirements discussed above. Moreover, it should reduce to the Kuhn kernel in appropriate limits in order to recover the observational data connected to the nearly flat rotation curves of spiral galaxies. However, these conditions are not sufficient to specify a unique functional form for $q$. 

Our physical considerations thus far involved the motion of stars and gas clouds in circular orbits around the galactic core. The radii of such orbits extend from the core radius to the outer reaches of the spiral galaxy. The resulting Kuhn kernel $q_K$ captures important physical aspects of the problem, but it is not mathematically suitable as it is not $L^1$ and $L^2$. In fact, $q_K$ integrated over all space leads to an infinite amount of effective dark matter for any point mass. The reciprocal kernel $q(\mathbf{r})$ of nonlocal gravity must satisfy the mathematical properties described above.  That is, from the standpoint of nonlocal gravity, the Tohline-Kuhn approach reflects the appropriate generalization of Newtonian gravity in the \emph{intermediate} galactic regime from the bulge to the outer limits of a spiral galaxy; however, the $r \to 0$ and $r \to \infty$ regimes are not taken into account. It follows from these considerations that $q$ must be constructed out of $q_K$ by moderating its short and long distance behaviors.

To proceed, let us start from the Kuhn kernel~\eqref{III2} and recall that it leads to flat rotation curves in the intermediate distance regime extending from the core radius to the outer limits of a spiral galaxy. The $r\to \infty$ behavior of $q$ is related to the fading of spatial memory with distance. If the decay rate of a quantity is proportional to itself, then the quantity dies out exponentially. We therefore adopt the simple rule that $q(r)$ behaves as $\exp{(-\mu_0\,r)}$ for $r\to \infty$, where $\mu_0^{-1}$ is a new length parameter that characterizes the rate of spatial decay of gravitational memory. For $r \ll \mu_0^{-1}$, where we expect to recover the nearly flat rotation curve of a spiral galaxy, the modified Kuhn kernel becomes
\begin{equation}\label{K1}
\frac{1}{4\pi \lambda}~ \frac{1}{r^2}~e^{-\mu_0\, r}= \frac{1}{4\pi \lambda}~ \frac{1}{r^2}\,(1-\mu_0\, r+\frac{1}{2}\, \mu_0^2\, r^2 -\cdots)\,,
\end{equation}
where the dominant correction is of linear order in $\mu_0\,r \ll 1$.  To cancel the linear correction in Eq.~\eqref{K1} and hence provide a better approximation to the Kuhn kernel for $\mu_0\,r \ll 1$, we consider instead
\begin{equation}\label{K2}
\frac{1}{4\pi \lambda}~ \frac{1}{r^2}~(1+\mu_0\,r)\,e^{-\mu_0\, r}= \frac{1}{4\pi \lambda}~ \frac{1}{r^2}\,\left[1-\frac{1}{2}\, (\mu_0\, r)^2 +\frac{1}{3}\, (\mu_0\, r)^3 -\cdots\right]\,.
\end{equation}

Kernel~\eqref{K2} is integrable over all space, but it is not square integrable. We must therefore modify the $r\to 0$ behavior of kernel~\eqref{K2} to make it square integrable by essentially replacing $r$ with $a_0+r$, where $a_0>0$ is a new constant length parameter. We note that two simple square-integrable  possibilities exist  
\begin{equation}\label{K3}
\frac{1}{4\pi \lambda}~ \frac{1+\mu_0 (a_0+r)}{r\,(a_0+r)}~e^{-\mu_0\,(a_0+ r)}\,
\end{equation}
and
\begin{equation}\label{K4}
\frac{1}{4\pi \lambda}~ \frac{1+\mu_0 (a_0+r)}{(a_0+r)^2}~e^{-\mu_0\,(a_0+ r)}\,.
\end{equation}
Moreover, we can  define  
\begin{equation}\label{K5}
\frac{1}{\lambda_0} := \frac{1}{\lambda}~e^{-\mu_0\,a_0}\,,
\end{equation}
so that the Tohline-Kuhn parameter $\lambda$ is modified and is henceforth replaced by $\lambda_0$. In this way, we find from Eqs.~\eqref{K3} and~\eqref{K4} two possible solutions for $q$, namely, $q_1$ and $q_2$ given by~\cite{NL5} 
\begin{equation}\label{IV1}
q_1=\frac{1}{4\pi \lambda_0}~ \frac{1+\mu_0 (a_0+r)}{r\,(a_0+r)}~e^{-\mu_0 r}\,
\end{equation}
and
\begin{equation}\label{IV2}
q_2=\frac{1}{4\pi \lambda_0}~ \frac{1+\mu_0 (a_0+r)}{(a_0+r)^2}~e^{-\mu_0 r}\,,
\end{equation}
where $r=|\mathbf{x}-\mathbf{y}|$ and $q_1$ and $q_2$ are symmetric functions of $\mathbf{x}$ and 
$\mathbf{y}$. Here, $\lambda_0$, $a_0$ and $\mu_0$ are three positive constant parameters that must be determined via observational data. The fundamental length scale of nonlocal gravity is 
$\lambda_0$, which is  expected to be of the order of 1\,kpc and is reminiscent of the parameter $\lambda$ of the Kuhn kernel.  We note that for $i=1,2$,  $q_i \to 0$ and nonlocality disappears as $\lambda_0 \to \infty$. Furthermore, $a_0$ moderates the $r \to 0$ behavior of the reciprocal kernel, while the kernel decays exponentially for $r\gg \mu_0^{-1}$, as the spatial gravitational memory fades.  Henceforth, we will refer to $a_0$ and $\mu_0$ as the short-distance and the large-distance parameters of the reciprocal kernel, respectively. 

In agreement with the requirements of the Fourier Transform Method,  kernels $q_1$ and $q_2$ are continuous positive functions that are integrable as well as square integrable over all space. The Fourier transform of $q_1$ is  always real and positive and hence satisfies Eq.~\eqref{II6} regardless of the value of $a_0/\lambda_0$. On the other hand, the Fourier transform of $q_2$ is such that Eq.~\eqref{II6} is satisfied if $a_0 < \lambda_0$.  In any case, it is natural to expect on physical grounds that $a_0<\lambda_0<\mu_0^{-1}$; that is, the (intermediate) nonlocality parameter is expected to be smaller than the large-distance parameter and larger than the short-distance parameter. It then follows from the Fourier Transform Method that the corresponding kernels $\chi_1$ and $\chi_2$ exist, are symmetric and have other desirable physical properties~\cite{NL5}.

It is important to emphasize that $q_1$ and $q_2$ are by no means unique. More complicated expressions that include more parameters are certainly possible. Kernels $q_1$ and $q_2$ appear to be the simplest functions that satisfy the requirements of nonlocal gravity theory discussed above~\cite{NL5}. 

The reciprocal kernels $q_1$ and $q_2$ thus depend upon three parameters: the nonlocality parameter $\lambda_0$, the large-distance parameter $\mu_0$ and the short-distance parameter $a_0$.  We expect that these three parameters will be determined via observational data, which will, in addition, point to a unique function (i.e., either $q_1$ or $q_2$) for $q$. 

It is interesting to note that for $a_0=0$, $q_1$ and $q_2$  both reduce to $q_0$,
\begin{equation}\label{IV3}
 q_0=\frac{1}{4 \pi \lambda_0}\frac{(1+\mu_0 r)}{r^2}e^{-\mu_0 r}\,,
\end{equation}
where for any finite $r: 0 \to \infty$, we have for $i=1,2$,
\begin{equation}\label{IV3a}
 q_0 (r) > q_i(r)\,.
\end{equation}
Moreover, $q_0$ is \emph{not} square integrable over all space and the behavior of $q_0$ for $r \to 0$ is precisely the same as that of the Kuhn kernel; for instance, in the Solar System, we recover the Tohline-Kuhn force~\eqref{III3}. For observational data related to the rotation curves of spiral galaxies as well as the internal  gravitational physics of clusters of galaxies, we expect that the short-distance behavior of the kernel would be unimportant and hence $q_0$ may be employed to fit the data. This has indeed been done in Ref.~\cite{NL6} and parameters $\lambda_0$ and $\mu_0$ have thus been determined. In this connection, it is useful to introduce the dimensionless parameter $\alpha_0$,
\begin{equation}\label{IV4}
 \alpha_0 :=\int q_0(|\mathbf{x}|)\,d^3x\,, \qquad \alpha_0 = \frac{2}{\lambda_0 \mu_0}\,.
\end{equation}
Then, it follows from observational data that~\cite{NL6} 
\begin{equation}\label{IV6}
\alpha_0 = 10.94\pm2.56\,, \qquad  \mu_0 = 0.059\pm0.028~{\rm kpc^{-1}}\,.
\end{equation}
Hence,  $\lambda_0=2/(\alpha_0\, \mu_0)$ turns out to be $\lambda_0 \approx 3 \pm 2~ {\rm kpc}$. \emph{It remains to determine $a_0$, $a_0<\lambda_0<\mu_0^{-1}$,  and hence the kernel (i.e., either $q_1$ or $q_2$) from observational data regarding the short-distance behavior of the reciprocal kernel.} To this end, it is useful to introduce a new parameter $p$, 
\begin{equation}\label{IV7}
p:= \mu_0\,a_0\,,
\end{equation}
and provisionally assume, on the basis of $a_0 < \lambda_0$ and Eq.~\eqref{IV6}, that 
\begin{equation}\label{IV8}
0< p < \frac{1}{5}\,
\end{equation}
for the sake of simplicity. 

It is abundantly clear from our considerations here that the choice of the kernel is not unique.  In the absence of a physical principle that could uniquely lead to the appropriate kernel, we  must adopt simple functional forms that satisfy the mathematical requirements discussed above and are based on agreement with observation. Let us recall that the relativistic framework of Einstein's field theory of gravitation has properly generalized Newton's  inverse square force law, which is ultimately based on Solar System observations that originally led to Kepler's laws of planetary motion. That is, an acceptable theory of gravitation must agree with Newton's theory in some form. How did Newton come up with the inverse square law? As explained in his \emph{Principia}, he explored various functional forms such as $r$ and $r^{-3}$ in addition to $r^{-2}$ and concluded that only $r^{-2}$ agreed with Kepler's empirical laws of planetary motion.  In short, the inverse square force law was not derived from a physical principle; rather, it was chosen to agree with observation. Moreover, observational data never have infinite accuracy; therefore, to Newton's $r^{-2}$, for example, one can add other functional forms with sufficiently small coefficients such that agreement with experimental results can be maintained. The same is true, of course, in Einstein's general  theory of relativity.

\section{Modified Force Laws}

In Ref.~\cite{NL6}, devoted to the astrophysical consequences of kernel $q_0$ defined in Eq.~\eqref{IV3}, the implications of the Tohline-Kuhn force for the Solar System were also discussed for the sake of completeness. In fact, parameter $a_0$ has been essentially ignored thus far in the interest of simplicity; this shortcoming is corrected in the present work. We now proceed to the determination of the short-distance behavior of the modified force laws associated with $q_1$ and $q_2$.

The gravitational force acting on a point particle of mass $m$ in a gravitational field with potential $\Phi$ is $\mathbf{F}=-m \nabla \Phi$ and the geodesic equation reduces in the Newtonian regime to Newton's equation of motion
\begin{equation}\label{V1}
\frac{d^2\mathbf{r}}{dt^2}=-\nabla \Phi(\mathbf{r})\,.
\end{equation}
Let us now imagine that potential $\Phi$ is due to a point mass $M$ at the origin of spatial coordinates with mass density  $\rho(\mathbf{r})=M\,\delta(\mathbf{r})$.  Thus we find from Eq.~\eqref{I3} that 
\begin{equation}\label{V2}
  \nabla^2\Phi_i (\mathbf{r}) = 4\pi G M \,[\delta (\mathbf{r})+  q_i(r)]\,,
\end{equation}
where $i=1, 2$,  depending upon which reciprocal kernel is employed, since experiment must ultimately decide between $q_1$ and $q_2$. Assuming that the force on a point mass $m$ at $\mathbf{r}$  due to $M$ is radial, namely, $\mathbf{F} = -m\, (d\Phi / dr)\, \hat{\mathbf{r}}$, where $\hat{\mathbf{r}}$ is the radial unit vector, we have
 \begin{equation}\label{V3}
 \frac{d\Phi}{dr}= G\, M \,f(r)\,,
\end{equation}
so that the gravitational force between the two point masses is $\mathbf{F} = -G\,m\, M\,f(r)\, \hat{\mathbf{r}}$.

The solution of Eq.~\eqref{V2} is the sum of the Newtonian potential plus $\phi_i(r)$, which is the contribution from the reciprocal kernel; that is, 
\begin{equation}\label{V3a}
\Phi_i(r) = G M \,\left[-\frac{1}{r}+\phi_i(r)\right]\,. 
\end{equation}
It follows from 
\begin{equation}\label{V4}
\nabla^2 \left(\frac{1}{r}\right) = - 4 \pi \delta (\mathbf{r})\,
\end{equation}
that 
\begin{equation}\label{V4a}
  \nabla^2\phi_i  = 4\pi\, q_i\,.
\end{equation}
It then proves useful to write
\begin{equation}\label{V5}
f_i(r) = \frac{1}{r^2} +N_i(r)\,, 
\end{equation}
where $N_i(r)=d\phi_i/dr$ and we have again separated the Newtonian contribution from the nonlocal contribution. Thus we find from Eqs.~\eqref{III1} and~\eqref{V4a} that
\begin{equation}\label{V6}
\frac{1}{r^2} \, \frac{d}{dr}[r^2 N_i(r)] = 4 \pi \,q_i(r)\,.
\end{equation}
The solution of this equation can be expressed as
\begin{equation}\label{V7}
N_i(r) = \frac{4\pi}{r^2} \, \int_0^r s^2\,q_i(s)\, ds\,,
\end{equation}
where we have assumed that  as $r \to 0$, $r^2\,N_i(r) \to 0$,  so that in the limit of $r \to 0$, the force on $m$ due to $M$ is given by the Newtonian inverse square force law. This important assumption is based on the results of  experiments  that have verified the gravitational inverse square force law down to a radius of $r \approx 50\, \mu$m~\cite{A1, A2, A3, A4}. Furthermore, no significant deviation from Newton's law of gravitation has been detected thus far in laboratory experiments~\cite{LL}. 

It proves interesting to define
\begin{equation}\label{V8}
N_0(r) :=   \frac{4\pi}{r^2} \, \int_0^r s^2\,q_0(s)\, ds = \frac{\alpha_0}{r^2}\,\left[ 1- (1+\frac{1}{2}\,\mu_0\,r)\,e^{-\mu_0\,r}\right]\,,
\end{equation}
where $q_0$ is given by Eq.~\eqref{IV3}, so that we can write 
\begin{equation}\label{V9}
N_i(r) = -\frac{1}{r^2}\,{\cal E}_i(r) +N_0(r)\,.
\end{equation}
Here, we have defined
\begin{equation}\label{V10}
{\cal E}_i(r) := 4\pi \, \int_0^r s^2\,[q_0(s)-q_i(s)]\, ds\,,
\end{equation}
such that ${\cal E}_i(r) =0$ for $a_0=0$  and ${\cal E}_i(r) > 0$ for $ r > 0$. It follows from Eq.~\eqref{V7} and the fact that  $q_1$ and $q_2$ are positive functions that $N_i(r) \ge 0$; therefore, $f_i(r) >0$ by Eq.~\eqref{V5}. Putting Eqs.~\eqref{V5}, \eqref{V8} and~\eqref{V9} together, we find
\begin{equation}\label{V11}
f_i(r) = \frac{1}{r^2}\,[1-{\cal E}_i(r)+ \alpha_0]\,-\frac{\alpha_0}{r^2}\,(1+\frac{1}{2}\,\mu_0\,r)\,e^{-\mu_0\,r}\,.
\end{equation}
Thus, we finally have the force of gravity on point mass $m$ due to point mass $M$, namely, 
\begin{equation}\label{V12}
\mathbf{F}_i(\mathbf{r}) = -GmM\,\frac{\hat{\mathbf{r}}}{r^2}\,\left\{[1-{\cal E}_i(r)+ \alpha_0] - \alpha_0\,(1+\frac{1}{2}\,\mu_0\,r)\,e^{-\mu_0\,r}\right\}\,,
\end{equation}
which, except for the ${\cal E}_i(r)$ term,  is due to kernel $q_0$. This force is conservative, satisfies Newton's third law of motion and is  \emph{always attractive}. The gravitational force of attraction in Eq.~\eqref{V12} consists of two parts: an enhanced attractive ``Newtonian" part and a repulsive ``Yukawa" part with an exponential decay length of $\mu_0^{-1} \approx 17$ kpc. The exponential decay in the  Yukawa term originates from the fading of spatial \emph{memory}. 

Imagine a uniform thin spherical shell of matter and a point mass $m$ inside the hollow shell. As is well known, Newton's inverse-square law of gravity implies that there is no net force on $m$, regardless of the location of $m$ within the shell. However, Newton's shell theorem does not hold in nonlocal gravity, so that $m$ would in general be subject to a gravitational force that is along the diameter that connects $m$ to the center of the shell and can be calculated by suitably  integrating 
$\mathbf{F}_i(\mathbf{r}) + GmM\,(1+\alpha_0)\,r^{-2}\,\hat{\mathbf{r}}$ over the shell, where $\mathbf{F}_i(\mathbf{r})$ is given by Eq.~\eqref{V12}. 

The short-distance parameter $a_0$ appears only in ${\cal E}_i(r)$; therefore, we now turn to the study of ${\cal E}_i(r)$. To this end, let us first define the \emph{exponential integral function}~\cite{A+S}
\begin{equation}\label{V13}
E_1(u):=\int_{u}^{\infty}\frac{e^{-t}}{t}dt\,.
\end{equation}
For $u: 0 \to \infty$,  $E_1(u)$ is a  positive function that monotonically decreases from infinity to zero. Indeed,  $E_1(u)$  behaves like $-\ln u$ near $u=0$ and vanishes exponentially as $u \to \infty$.  Moreover, 
\begin{equation}\label{V14}
E_1(x)=-C-\ln x -\sum_{n=1}^{\infty}\frac{(-x)^n}{n~ n!}\,,
\end{equation}
where $C=0.577\dots$ is Euler's constant. It is useful to note that 
\begin{equation}\label{V15}
 \frac{e^{-u}}{u+1} < E_1(u) \le \frac{e^{-u}}{u}\,,
\end{equation}
see formula 5.1.19 in Ref.~\cite{A+S}. 

From Eq.~\eqref{V10}, we find by straightforward integration that 
\begin{equation}\label{V16}
 {\cal E}_1(r)=\frac{a_0}{\lambda_0}e^{p}\Big[E_1(p)-E_1(p+\mu_0 r)\Big]\,
\end{equation}
and
\begin{equation}\label{V17}
 {\cal E}_2(r)=\frac{a_0}{\lambda_0}\left \{-\frac{r}{r+a_0}e^{-\mu_0 r}+2e^{p} \Big[E_1(p)-E_1(p+\mu_0 r) \Big] \right \}\,,
\end{equation}
where $p$ has been defined in Eqs.~\eqref{IV7} and~\eqref{IV8}.
Furthermore, it follows from Eq.~\eqref{V10} that 
\begin{equation}\label{V18}
\frac{d{\cal E}_i}{dr} = 4\pi \, r^2\,[q_0(r)-q_i(r)]\,,
\end{equation}
where the right-hand side is positive by Eq.~\eqref{IV3a}. More explicitly,
\begin{equation}\label{V19}
\frac{d{\cal E}_1}{dr} = \frac{a_0}{\lambda_0}\, \frac{1}{a_0+r}\,e^{-\mu_0\,r}\,
\end{equation}
and
\begin{equation}\label{V20}
\frac{d{\cal E}_2}{dr} = \frac{a_0}{\lambda_0}\,\left[\mu_0 + \frac{2-p}{a_0+r}-\frac{a_0}{(a_0+r)^2}\right]\,e^{-\mu_0\,r}\,.
\end{equation}
Thus ${\cal E}_1(r)$ and $ {\cal E}_2(r)$ are positive, monotonically increasing functions of $r$ that start from zero at $r=0$ and asymptotically approach,  for $r \to \infty$,  ${\cal E}_1(\infty)={\cal E}_{\infty}$  and ${\cal E}_2(\infty)= 2\,{\cal E}_{\infty}$, respectively. Here,  
\begin{equation}\label{V21}
 {\cal E}_{\infty}=\frac{1}{2}\,\alpha_0\, p\,e^{p}E_1(p)\,.
\end{equation}
It then follows from Eq.~\eqref{V15} that 
\begin{equation}\label{V22}
 {\cal E}_{\infty} < \frac{\alpha_0}{2}\,,
\end{equation}
so that in formula~\eqref{V12} for the gravitational force, 
\begin{equation}\label{V22}
\alpha_0 -  {\cal E}_i(r) >0\,.
\end{equation}
Thus for $r \gg \mu_0^{-1}$, the Yukawa part of Eq.~\eqref{V12} can be neglected and
\begin{equation}\label{V22a}
\mathbf{F}_i(\mathbf{r}) \approx -\frac{GmM\,[1+ \alpha_0-{\cal E}_i(\infty)] }{r^2}\,\hat{\mathbf{r}}\,,
\end{equation} 
so that $M\,[\alpha_0-{\cal E}_i(\infty)]$ has the interpretation of the total effective dark mass associated with $M$.

For $a_0=0$, the net effective dark matter associated with point mass $M$ is simply $\alpha_0\,M$, where $\alpha_0 \approx 11$. On the other hand, for $a_0 \ne 0$, the corresponding result is $\alpha_0\, \epsilon_i(p)\, M$, where
\begin{equation}\label{V22aA}
\epsilon_1(p) = 1-\frac{1}{2}\,p\,e^p\,E_1(p)\,, \qquad \epsilon_2(p) = 1-p\,e^p\,E_1(p)\,. 
\end{equation}
These functions are plotted in Figure 1 for $p: 0 \to 0.2$ in accordance with Eq.~\eqref{IV8}. 

\begin{figure}
\centerline{\includegraphics[width=4.2in]{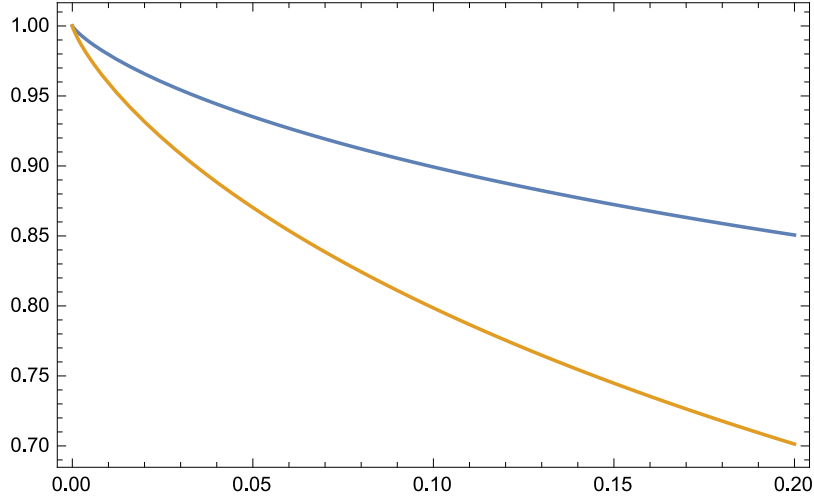}}
\caption{The figure depicts the graph of the function $\epsilon_1(p)$ that lies above the graph of $\epsilon_2(p)$ for $0<p<1/5$.}
\end{figure}

Finally, let us note that the solution of Eq.~\eqref{V3} for the gravitational potential $\Phi_i$ due to a point mass $M$ at $r=0$ is given by
\begin{equation}\label{V22b}
\Phi_i(r)=GM\int_{\infty}^r f_i(r') dr'\,,
\end{equation}
where, as expected, we have assumed that $\Phi_i(r) \to 0$, when $r \to \infty$. It follows from a detailed but straightforward calculation that for $i=1, 2$, corresponding to $q_1$ and $q_2$, respectively, 
\begin{equation}\label{V22c}
 \Phi_1(r) = -\frac{GM}{r}\,\left(1+\alpha_0 -\, {\cal E}_{\infty}-\alpha_0\,e^{-\mu_0\,r}\right) 
-\frac{GM}{\lambda_0}\,(1+\frac{a_0}{r})\,e^{p}\,E_1(p+\mu_0\,r)\,
\end{equation}
and 
\begin{equation}\label{V22d}
 \Phi_2(r) = -\frac{GM}{r}\,\left(1+\alpha_0 -2\, {\cal E}_{\infty}-\alpha_0\,e^{-\mu_0\,r}\right) 
-\frac{GM}{\lambda_0}\,(1+2\,\frac{a_0}{r})\,e^{p}\,E_1(p+\mu_0\,r)\,.
\end{equation}
In these expressions, we can use Taylor expansion of $E_1(p+\mu_0\,r)$ about $p=\mu_0\,a_0$ to write
\begin{equation}\label{V22e}
e^{p}\,E_1(p+\mu_0\,r)=\frac{\lambda_0}{a_0}\,{\cal E}_{\infty}-\frac{r}{a_0}+ \frac{1}{2}\,(1+p)\,\frac{r^2}{a_0^2} -\cdots\,.
\end{equation}
In this way, we see that  $\Phi_i(r) \to -GM/r$ for $r\to 0$.  It follows from Eqs.~\eqref{V22c} and~\eqref{V22d} that in the limiting case where $a_0=0$, we have $\Phi_1=\Phi_2=\Phi_0$, where 
\begin{equation}\label{V22f}
 \Phi_0(r) = -\frac{GM}{r}\,\left(1+\alpha_0 -\alpha_0\,e^{-\mu_0\,r}\right) 
-\frac{GM}{\lambda_0}\,E_1(\mu_0\,r)\,
\end{equation}
is the gravitational potential corresponding to kernel $q_0$.

\subsection{Short-Distance Behavior of the Gravitational Force}

It is natural to assume that the short-distance parameter $a_0$,  $a_0<\lambda_0<\mu_0^{-1}$,  may eventually turn out to be much smaller than the nonlocality parameter $\lambda_0$.   For instance, with $a_0/\lambda_0=10^{-3}$ and the parameters of our nonlocal gravity model as in Eq.~\eqref{IV6}, we have ${\cal E}_{\infty}\approx 0.008$. Thus if $a_0\ll \lambda_0$, then  in such a case, $0<  {\cal E}_{\infty} \ll 1$ and for most astrophysical applications ${\cal E}_i(r)$ in the force law~\eqref{V12} may simply be neglected in comparison to unity~\cite{NL6}. However, ${\cal E}_i(r)$ is crucial for the discussion of the short-distance behavior of the gravitational force. To investigate this point, let us first find  the Taylor expansion of ${\cal E}_i(r)$ about $r=0$. 
From Eqs.~\eqref{V19} and~\eqref{V20}, it is straightforward to show by repeated differentiation that 
\begin{equation}\label{V23}
{\cal E}_1(r) = \frac{r}{\lambda_0}\,\left[1-\frac{1}{2}\, W_1(p)\,\left(\frac{r}{a_0}\right)+\frac{1}{3}\, W_2(p)\,\left(\frac{r}{a_0}\right)^2 - \cdots\right]\,
\end{equation}
and
\begin{equation}\label{V24}
{\cal E}_2(r) = \frac{r}{\lambda_0}\,\left[1-\frac{1}{3}\, W_2(p)\,\left(\frac{r}{a_0}\right)^2 + \cdots\right]\,,
\end{equation}
where
\begin{equation}\label{V25}
W_1(p)=1+p\,, \qquad W_2(p) = 1+p+\frac{1}{2}\,p^2\,.
\end{equation}
Thus, we find from Eq.~\eqref{V12} that 
\begin{equation}\label{V26}
\mathbf{F}_1(\mathbf{r}) = -GmM\,\frac{\hat{\mathbf{r}}}{r^2}\,\left[1+\frac{1}{2}\,(1+p)\,\frac{r^2}{\lambda_0\,a_0}- \frac{1}{3}\, (1+ p +  p^2 )\,\frac{r^3}{\lambda_0\,a_0^2} + \cdots\right]\,
\end{equation}
and
\begin{equation}\label{V27}
\mathbf{F}_2(\mathbf{r}) = -GmM\,\frac{\hat{\mathbf{r}}}{r^2}\,\left[1+\frac{1}{3}\, (1+ p)\,\frac{r^3}{\lambda_0\,a_0^2} + \cdots\right]\,.
\end{equation}
It is remarkable that  in the square brackets in Eqs.~\eqref{V26} and~\eqref{V27}, the linear $r/\lambda_0$ term is absent; in fact, this is the leading term in both ${\cal E}_1(r)$ and ${\cal E}_2(r)$, but is simply canceled by the corresponding Tohline-Kuhn term coming from $q_0$. Thus it appears that the existence of $a_0\ne 0$ in effect shields the near-field region from the influence of the $1/r$ part of the Tohline-Kuhn force.

It follows from these results that the main nonlocal deviation  from the Newtonian inverse square force law in the two-body system, $\delta \mathbf{F}$, could be either of the form
\begin{equation}\label{V28}
\delta \mathbf{F}_1(\mathbf{r}) = -\frac{1}{2} \, \frac{GmM}{\lambda_0\,a_0}\,(1+ p)\,\hat{\mathbf{r}}+ \frac{1}{3} \, \frac{GmM}{\lambda_0\,a_0}\,(1+ p+ p^2 )\,\frac{r}{a_0}\,\hat{\mathbf{r}}\,
\end{equation}
if kernel $q_1$ is employed, or 
\begin{equation}\label{V29}
\delta \mathbf{F}_2(\mathbf{r}) = -\frac{1}{3} \, \frac{GmM}{\lambda_0\,a_0}\,(1+ p)\,\frac{r}{a_0}\,\hat{\mathbf{r}}\,
\end{equation}
if kernel $q_2$ is employed. Here, $a_0<\lambda_0<\mu_0^{-1}$; indeed, let us note that with $\lambda_0 \approx 3$ kpc and $\mu_0^{-1} \approx 17$ kpc, we expect that $p=\mu_0\, a_0$, $0<p<1/5$, would be rather small in comparison with unity. 

\section {Kepler System}

Imagine a Keplerian two-body system of point particles with a \emph{radial} perturbing acceleration 
$\boldsymbol{\mathcal{A}}=\delta \mathbf{F}/m$, 
\begin{equation}\label{3.1}
\frac{d^2\mathbf{r}}{dt^2} + \frac{GM\mathbf{r}}{r^3}=\boldsymbol{\mathcal{A}}\,.
\end{equation}
The orbital angular momentum of the system is then conserved and the orbit remains planar. Consider first the case where the radial acceleration is of the form $\boldsymbol{\mathcal{A}}= \eta\, \mathbf{r}$, where $\eta$ is a constant. It can be shown using the Lagrange planetary equations, when averaged over the fast Keplerian motion with orbital frequency $\omega_0$, $\omega_0^2=GM/A^3$, that the orbit keeps its shape but slowly precesses. That is, the semimajor axis of the orbit $A$ and the orbital eccentricity $e$ remain constant on the average, but there is a slow pericenter precession whose frequency is given by $\Omega \, \hat{\boldsymbol{\ell}}$,
 where~\cite{KHM}
\begin{equation}\label{3.2}
\Omega = \frac{3}{2}\,\frac{\eta}{\omega_0}\, \sqrt{1-e^2}\,
\end{equation}
and $\hat{\boldsymbol{\ell}}$ is the unit orbital angular momentum vector. 

This case is reminiscent of the orbital perturbation due to the presence of a cosmological constant~\cite{KHM}. Moreover, Eq.~\eqref{3.2} can also be obtained from the study of the average precession of the Runge-Lenz vector due to the presence of the perturbing acceleration~\cite{KHM}. 

Similarly, if the perturbing acceleration is radial and \emph{constant}, namely, $\boldsymbol{\mathcal{A}}= \eta'\, \hat{\mathbf{r}}$, then, as before,  the shape of the orbit remains constant on the average, but there is a slow pericenter precession of frequency $\Omega' \, \hat{\boldsymbol{\ell}}$, where
\begin{equation}\label{3.3}
\Omega' = \frac{\eta'}{\omega_0\, A}\, \sqrt{1-e^2}\,.
\end{equation}
This result has been noted before in connection with studies of the Pioneer anomaly~\cite{Iorio2, Sa, SJ}.

It follows from the results of the previous section that in nonlocal gravity the orbit on average remains planar and keeps its shape, but slowly precesses.  If the reciprocal kernel of nonlocal  gravity in the Newtonian regime is $q_1$,  then $\delta \mathbf{F}_1(\mathbf{r})=m\,(\eta'+ \eta_1\, r)\,\hat{\mathbf{r}}$, where
\begin{equation}\label{3.4}
\eta' = -\frac{1}{2} \,\frac{GM}{\lambda_0\,a_0}\,(1+p)\,, \quad \eta_1= \frac{1}{3} \,\frac{GM}{\lambda_0\,a_0^2}\,(1+p + p^2)\,.
\end{equation}
Thus, superposing small perturbations, we get for the pericenter advance in this case that 
\begin{equation}\label{3.5}
\Omega_1 = -\frac{1}{2} \,\omega_0\,\frac{A^2}{\lambda_0\,a_0}\,\left[1+p-\frac{A}{a_0}\,(1+p + p^2)\right]\, \sqrt{1-e^2}\,.
\end{equation}

On the other hand, if the reciprocal kernel turns out to be $q_2$, then $\delta \mathbf{F}_2(\mathbf{r})=m\,\eta_2\,\mathbf{r}$, where
\begin{equation}\label{3.6}
\eta_2 = -\frac{1}{3} \frac{GM}{\lambda_0\,a_0^2}\,(1+p)\, 
\end{equation}
and hence the rate of advance of pericenter is \emph{negative} and is given by
\begin{equation}\label{3.7}
\Omega_2 = -\frac{1}{2} \,\omega_0\,\frac{A^3}{\lambda_0\,a_0^2}\,(1+p)\, \sqrt{1-e^2}\,.
\end{equation}

It is interesting to explore the implications of these results for the Solar System. This is the subject of the next section. 

\section{Perihelion Precession}

Thus far we have dealt with the force between point particles. To apply our results to realistic systems, such as the core of galaxies, binary pulsars or the Solar System, we need to investigate the influence of the finite size of an astronomical body on the attractive gravitational force that it can generate. To simplify matters, imagine a point mass $m$ outside a spherically symmetric body of  radius $R_0$ that has  uniform density  and total mass $M$. Let $R$ be the distance between $m$ and the center of the sphere, so that $R>R_0$.  If the force of gravity is radial, we expect by symmetry that the net force on $m$ would be along the line joining the center of the sphere to $m$. Under what conditions would the spherical body act on $m$ as though its mass were concentrated at its center? It turns out that, in addition to Newton's law of gravity, any radial force that is proportional to distance would work just as well, so that in general the desired two-body force can be any linear superposition of these forces such as in the case of kernel $q_2$ and Eq.~\eqref{V27}.  On the other hand, in connection with kernel $q_1$ and Eq.~\eqref{V26}, we find, after a detailed but straightforward calculation,  that for a \emph{constant} radial force the same is true, except that the strength of the constant force is thereby reduced by a factor of
\begin{equation}\label{4.1}
1-\frac{1}{5}\left(\frac{R_0}{R}\right)^2\,.
\end{equation}
This factor is nearly unity in most applications of interest here and we therefore assume that we can treat  uniform spherical bodies like point particles for the sake of simplicity.  This means that we can approximately  apply the results of the previous section to the influence of the Sun on the motion of a planet in the Solar System.

The recent advances in the study of precession of perihelia of planetary orbits have been reviewed by Iorio~\cite{Iorio}. In absolute magnitude, for instance, the extra perihelion shift of Mercury and Saturn due to nonlocal gravity would be expected to be less than about 10 and 2 milliarcseconds per century, respectively; otherwise, the effect of nonlocality would have already shown up in high-precision ephemerides~\cite{FLK, PP}, barring certain exceptional circumstances. Thus if the kernel of nonlocal gravity is $q_1$, the nonlocal contribution to the perihelion precession $\Omega_1$ is expected to  be
such that its absolute magnitude for Mercury and Saturn would be less than about $10^{-2}$ and $2\times 10^{-3}$ seconds of arc per century, respectively. In general, the inequality involving $|\Omega_i|$ under consideration here for $q_i$, $i=1,2$, gives a lower limit on $a_0$ that increases with $A$ as $A^{1/2}$ or $A^{3/4}$ depending on whether we choose $q_1$ or $q_2$, respectively. Thus the lower limit on $a_0$ can become more significant the farther the planetary orbit is from the Sun. 

For the orbit of Mercury, $A\approx 6\times 10^{12}$ cm and $e\approx 0.2$; moreover, the orbital period is about $0.24$ yr. If the reciprocal kernel is $q_1$, it follows  from Eq.~\eqref{3.5} and $\lambda_0 \approx 3$ kpc that in this case, $a_0 \gtrsim  7 \times10^{13}$ cm. Similarly, if the kernel is $q_2$, we find from Eq.~\eqref{3.7} that in this case, $a_0 \gtrsim 2 \times 10^{13}$ cm. 

For the orbit of Saturn, the orbital period is about 29.5 yr, $A \approx 1.4 \times 10^{14}$ cm and $e\approx 0.056$.  In a similar way, it follows that if the reciprocal kernel is $q_1$, $a_0 \gtrsim 2 \times 10^{15}$ cm. However, if the kernel is $q_2$, then $a_0 \gtrsim  5.5 \times10^{14}$ cm.

These preliminary lower limits can be significantly strengthened if, in the analysis of planetary data, Newton's law of gravity is replaced by either $\mathbf{F}_1$ given in Eq.~\eqref{V26} or $\mathbf{F}_2$ given in Eq.~\eqref{V27}, depending upon whether the reciprocal kernel of nonlocal gravity is chosen to be $q_1$ or $q_2$, respectively. In fact, nonlocal gravity in the Solar System could be tested experimentally via ESA's Gaia mission, launched in 2013, or other possible missions dedicated to measuring deviations from Newtonian gravity in the Solar System~\cite{HHP, BDG}.

\section{Gravitational Deflection of Light}

Light rays follow null geodesics in nonlocal gravity~\cite{NL7}. Consider the propagation of a light ray with impact parameter $\zeta$ in the gravitational field generated by a point mass $M$ that is essentially fixed at $r=0$. It is well known that in the linear post-Newtonian approximation, the total deflection angle of the light ray is twice the Newtonian expectation~\cite{NL3, NL6}. Therefore, if $\Delta$ is the net deflection angle, we have for $i=1,2$, 
\begin{equation}\label{L1}
 \Delta_i=\frac{4GM\,\zeta}{c^2}\int_0^{\frac{\pi}{2}}f_i\Big(\frac{\zeta}{\sin \vartheta}\Big)~\frac{d\vartheta}{\sin \vartheta}\,,
\end{equation}
where $f_i(r)$ is given by Eq.~\eqref{V11}. Here, $\zeta=r \sin \vartheta$ is the impact parameter and  
$\vartheta: 0 \to \pi$ is the corresponding scattering angle~\cite{NL6}.   

For $a_0=0$,  the reciprocal  kernel is then $q_0$ and the net deflection angle $\Delta_0$ has been studied in some detail in Refs.~\cite{NL3, NL6}. For our present purposes, $\Delta_0$ can be expressed as 
\begin{equation}\label{L2}
\Delta_0=\frac{4GM}{c^2\,\zeta}\,[1+\alpha_0\, {\cal I}_0(h)]\,,
\end{equation}
where  
\begin{equation}\label{L3}
h := \mu_0\, \zeta\,, \qquad  {\cal I}_0(h) :=1-\int_0^{\frac{\pi}{2}}(\cos \varphi +\frac{1}{2}\, h)\,e^{- h\,\sec \varphi}\,d\varphi\,
\end{equation}
and $\varphi +\vartheta=\pi/2$. For dimensionless impact parameter $h: 0 \to \infty$, we note that ${\cal I}_0(h): 0 \to 1$; that is, ${\cal I}_0(h)$ monotonically increases from zero and asymptotically approaches unity as $h \to \infty$. For $0<h \ll 1$, ${\cal I}_0(h)\approx \pi h/4$ and hence $\Delta_0$ differs from the Einstein deflection angle $\Delta_E=4GM/(c^2\, \zeta)$ by a constant angle that is proportional to the mass of the source and coincides with the result derived from the Tohline-Kuhn force law~\cite{NL3, NL6}. It is indeed smaller than the Einstein deflection angle $\Delta_E$ by a factor of $\sim 10^{-11}$ for light rays passing near the rim of the Sun. 

In nonlocal gravity,  $a_0 > 0$ and we find from Eqs.~\eqref{V11} and~\eqref{L1} that 
\begin{equation}\label{L4}
\Delta_i=\Delta_0 - \frac{4GM}{c^2\, \zeta}\,\int_0^{\frac{\pi}{2}} \cos \varphi ~ {\cal E}_i\Big(\frac{\zeta}{\cos \varphi}\Big)\,d\varphi\,.
\end{equation}
We can therefore write
\begin{equation}\label{L5}
\Delta_i=\Delta_E\,[1+\alpha_0\, {\cal I}_0(h) - {\cal E}_i(\infty)\,{\cal J}_i(h, p)]\,,
\end{equation}
where 
\begin{equation}\label{L6}
{\cal J}_i(h, p) :=\frac{1}{{\cal E}_i(\infty)}\int_0^{\frac{\pi}{2}} \cos \varphi ~ {\cal E}_i\Big(\frac{\zeta}{\cos \varphi}\Big)\,d\varphi\,.
\end{equation}
The functions ${\cal E}_1(r)$ and ${\cal E}_2(r)$ are given in Eqs.~\eqref{V16} and~\eqref{V17}, respectively. It turns out that for $r/a_0 \ll 1$, ${\cal E}_i(r) \approx r/\lambda_0$ by Eqs.~\eqref{V23} and~\eqref{V24}; hence, for $0<h \ll 1$, ${\cal E}_i(\infty)\,{\cal J}_i(h, p)\approx \alpha_0(\pi h/4)$. As expected, this term cancels the other (Tohline-Kuhn) term, $\alpha_0\, {\cal I}_0(h)\approx \alpha_0(\pi h/4)$, in Eq.~\eqref{L5}. Moreover, for $h: 0 \to \infty$, we note that ${\cal J}_i(h, p): 0 \to 1$; that is, ${\cal J}_i(h, p)$ monotonically increases from zero at $h=0$ and asymptotically approaches unity as $h \to \infty$. Thus for $h \gg 1$, i.e., large impact parameters  $\zeta \gg \mu_0^{-1}$, $\Delta_i \to \Delta_E [1+\alpha_0 - {\cal E}_i(\infty)]$, which is consistent with Eq.~\eqref{V22a}.  Indeed, we recall from Eq.~\eqref{V22aA} that 
$\alpha_0 - {\cal E}_i(\infty)=\alpha_0\, \epsilon_i(p)$, see Figure 1. That is, the extra deflection angle takes due account of the effective dark matter associated with $M$. 

\begin{figure}
\centerline{\includegraphics[width=4.2in]{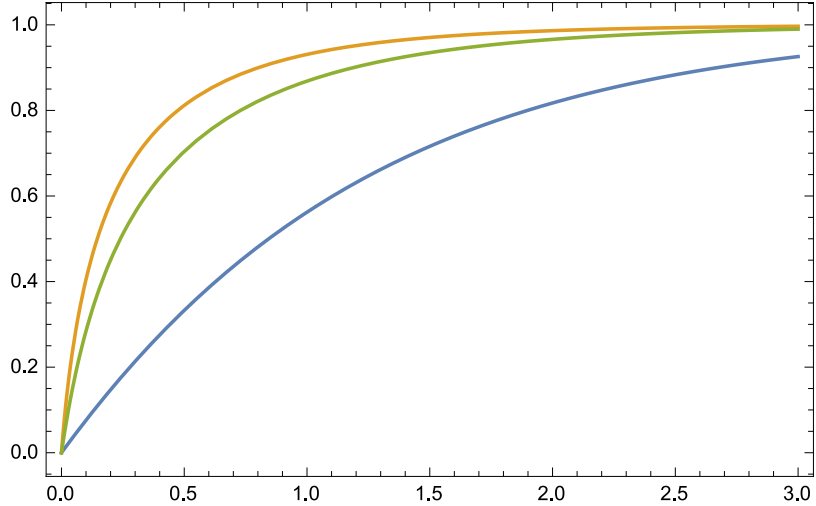}}
\caption{The figure depicts the graphs of the functions ${\cal I}_0(h)$, ${\cal J}_1(h, p)$ and   ${\cal J}_2(h, p)$ for $p=0.1$ and $h: 0 \to \infty$. The graph of  ${\cal J}_1$ lies above the graph of  ${\cal J}_2$, which in turn lies above the graph of  ${\cal I}_0$. }
\end{figure}

The new integral, ${\cal J}_i(h, p)$, can be expressed in terms of $d{\cal E}_i/dr$. Using integration by parts, Eq.~\eqref{L6} can be written as
\begin{equation}\label{L7}
{\cal J}_i(h, p) = 1 - \frac{\zeta}{{\cal E}_i(\infty)}\int_0^{\frac{\pi}{2}} \tan^2 \varphi ~ \frac{d{\cal E}_i}{dr}\Big(\frac{\zeta}{\cos \varphi}\Big)\,d\varphi\,,
\end{equation}
where  $d{\cal E}_1/dr$ and  $d{\cal E}_2/dr$ are given by Eqs.~\eqref{V19} and~\eqref{V20}, respectively. More explicitly, we have 
\begin{equation}\label{L8}
{\cal J}_1(h, p) = 1 - \frac{h}{e^p\,E_1(p)}\int_0^{\frac{\pi}{2}} \frac{\sin^2 \varphi}{(h+p\,\cos \varphi)\,\cos \varphi} ~ e^{- h\,\sec \varphi}\,d\varphi\,
\end{equation}
and
\begin{equation}\label{L9}
{\cal J}_2(h, p) = {\cal J}_1(h, p) - \frac{1}{2}\,\frac{h}{e^p\,E_1(p)}\int_0^{\frac{\pi}{2}} \frac{\cos \varphi}{h+p\,\cos \varphi} ~ e^{- h\,\sec \varphi}\,d\varphi\,.
\end{equation}
We plot ${\cal I}_0(h)$, ${\cal J}_1(h, p)$ and ${\cal J}_2(h, p)$ for $p=0.1$ and $h: 0 \to \infty$ in Figure 2.

It is possible to express the net deflection angle as
\begin{equation}\label{L10}
\Delta_i=\Delta_E\,[1+\alpha_0\, \Sigma_i(h, p)]\,,
\end{equation}
where $\Sigma_1$ and $\Sigma_2$ are given by
\begin{equation}\label{L11}
\Sigma_1(h, p) = {\cal I}_0(h) - \frac{1}{2}\,p\,e^p\,E_1(p)\,{\cal J}_1(h, p)\,
\end{equation}
and
\begin{equation}\label{L12}
\Sigma_2(h, p) = {\cal I}_0(h) - p\,e^p\,E_1(p)\,{\cal J}_2(h, p)\,,
\end{equation}
respectively.

It now remains to discuss the influence of $a_0>0$ on the gravitational deflection of starlight by the Sun. If the reciprocal kernel is $q_i, i=1,2$, then the net deflection angle due to nonlocality is $\alpha_0\, \Sigma_i(h, p)$ times the Einstein deflection angle $\Delta_E$, in accordance with Eq.~\eqref{L10}. For light rays passing near the rim of the Sun, the dimensionless impact parameter is very small ($h=10^{-12}$). Moreover, using the lower limits placed on $a_0$ in the previous section, we note that $p:\, \sim 4\times 10^{-8} \to 0.2$\, in $\Sigma_1$, while $p:\,\sim 10^{-8} \to 0.2$\, in $\Sigma_2$. Our numerical results indicate that $|\Sigma_1|$ and $|\Sigma_2|$ are negligibly small compared to unity. For instance, for $h=10^{-12}$ we find both $\Sigma_1$ and $\Sigma_2$ to be  $ \approx -10^{-15}$. To illustrate the situation, we plot $\Sigma_1$ and $\Sigma_2$ in Figures 3 and 4,  respectively. 

It is important to point out that ${\cal I}_0(h)$ and  ${\cal J}_i(h, p)$ are not analytic at $h=0$, so that they cannot be expanded in a Taylor series about $h=0$. The behavior of these functions for $h \to 0$ can in principle be determined using asymptotic approximation methods~\cite{BH}. A simple case is illustrated in the Appendix.

\begin{figure}
\centerline{\includegraphics[width=4.2in]{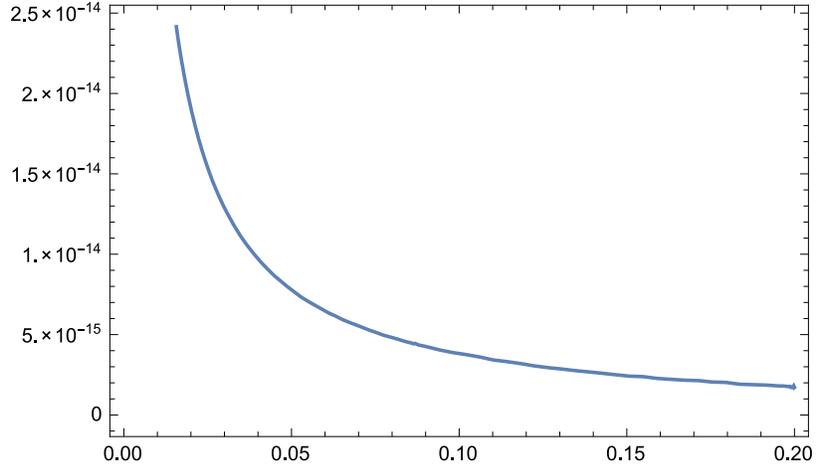}}
\caption{The figure depicts the graph of the function $p\mapsto \Sigma_1(10^{-8},p)$. We note that for  $p \approx  4 \times 10^{-8}$, $\Sigma_1 \approx 1.6 \times 10^{-9}$ in this case.}
\end{figure}

\begin{figure}
\centerline{\includegraphics[width=4.2in]{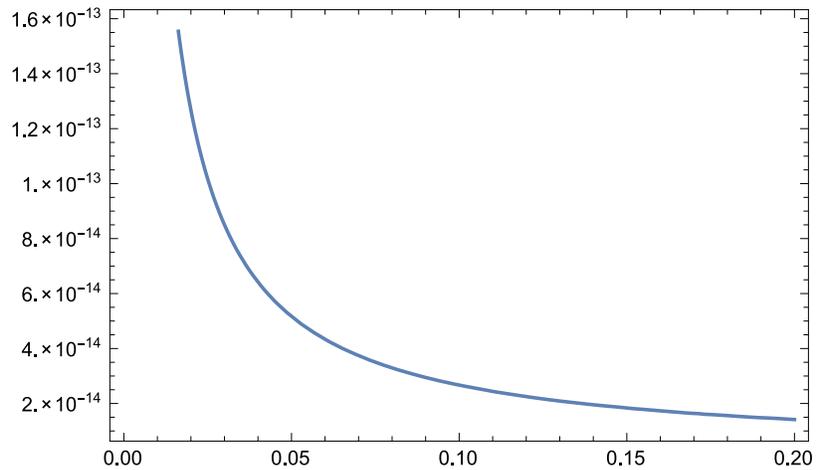}}
\caption{The figure depicts the graph of the function $p\mapsto \Sigma_2 (10^{-7},p)$. We note that for  $p \approx 10^{-8}$, $\Sigma_2 \approx 5.6 \times 10^{-8}$ in this case.}

\end{figure}

\section{Gravitational Time Delay}

The general expressions for the gravitational potentials corresponding to the reciprocal kernels $q_1$ and $q_2$ are given in Eqs.~\eqref{V22c} and~\eqref{V22d}, respectively. Within the Solar System, $\mu_0\,r \ll 1$ and we can therefore use expansions in powers of this small quantity as in Eq.~\eqref{V22e}. Neglecting terms of order $(r/a_0)^2$ and higher, we find
\begin{equation}\label{D1}
 \Phi_1(r) \approx -\frac{GM}{r}-\frac{GM}{\lambda_0}\,[1+e^{p}\,E_1(p)]+\frac{1}{2}\,\frac{GM}{\lambda_0}\,(1+p)\,\frac{r}{a_0}\,
\end{equation}
and 
\begin{equation}\label{D2}
 \Phi_2(r) \approx -\frac{GM}{r}-\frac{GM}{\lambda_0}\,e^{p}\,E_1(p)\,.
\end{equation}
The nonlocal contribution to the gravitational potential is extremely small within the Solar System. To illustrate this point, consider, for instance,  the gravitational shift of the frequency of light, which involves the difference  in the potential at two spatially separated events.  In the approximation scheme under consideration here, the contribution to the shift in the potential due to nonlocality is nonzero only in the case of $\Phi_1$ and is given by
\begin{equation}\label{D3}
\frac{1}{2}\,\frac{GM}{\lambda_0}\,(1+p)\,\frac{r_2-r_1}{a_0}\,,
\end{equation}
 where $r_1$ and $r_2$ are the radial positions of the events under consideration. This is rather small in absolute magnitude when compared with the corresponding shift of the Newtonian potential.  That is, at a distance of ${\cal L}=10$ astronomical units, say, we have ${\cal L}/\lambda_0 \sim 10^{-8}$ and 
${\cal L}/a_0 < 10^{-1}$ based on the lower limit on $a_0$ established in section IV. Therefore, we conclude that the relative contribution of nonlocality to the gravitational shift of the frequency of light is very small within the Solar System. 

Consider next the gravitational time delay ${\cal D}$ of a light signal that travels from event $P_1:(ct_1,\mathbf{r}_1)$ to event $P_2:(ct_2,\mathbf{r}_2)$. Then, 
${\cal D}=t_2-t_1-|\mathbf{r}_2-\mathbf{r}_1|/c$ is given by
\begin{equation}\label{D4}
{\cal D}_i =-\frac{2}{c^3}\int_{P_1}^{P_2}\Phi_i \,  dL\,,
\end{equation}
where $L: 0 \to |\mathbf{r}_2-\mathbf{r}_1|$ is the distance along a straight line from $P_1$ to $P_2$.
It is in general straightforward to compute ${\cal D}_i$ for nonlocal gravity in the Solar System. However, to simplify matters, we consider only the time delay due to  $\Phi_2$, which is  
\begin{equation} \label{D5}
{\cal D}_2= \frac{2GM}{c^3}\,\left[\, \ln
  \frac{r_2+\hat{\mathbf{n}}\cdot \mathbf{r}_2}
  {r_1+\hat{\mathbf{n}}\cdot \mathbf{r}_1}  +e^{p}\,E_1(p)\, \frac{|\mathbf{r}_2-\mathbf{r}_1|}{\lambda_0}\right]\,,
\end{equation}
where  $\hat{\mathbf{n}}= (\mathbf{r}_2-\mathbf{r}_1)/|\mathbf{r}_2-\mathbf{r}_1|$. The result is simply the sum of the Shapiro time delay and the nonlocal contribution to signal retardation.   We recall that $p: \,\sim 10^{-8} \to 0.2$\, in this case; moreover, it follows from Eq.~\eqref{V14} that for $0<p \ll 1$, $E_1(p) \approx -C-\ln p$. If  $|\mathbf{r}_2-\mathbf{r}_1|$ is about an astronomical unit, then $|\mathbf{r}_2-\mathbf{r}_1|/\lambda_0 \sim 10^{-9}$; therefore, the nonlocal effect is rather small and probably  difficult to measure, since there are uncertainties due to clock stability as well as the existence of the interplanetary medium~\cite{Shap}.

\section{Discussion}

The Newtonian regime of nonlocal gravity involves a modified Poisson equation with a reciprocal kernel $q$. Two possible functional forms for $q$, namely,  $q_1$ and $q_2$, have been explicitly determined on the basis of a detailed investigation~\cite{NL5}. Each such kernel contains three parameters that all have dimensions of length: $\lambda_0$, $\mu_0^{-1}$ and $a_0$. Furthermore, we have  $a_0<\lambda_0<\mu_0^{-1}$. For $a_0=0$, there is much simplification, since $q_1=q_2=q_0$, where the parameters of kernel $q_0$, namely the basic nonlocality length scale $\lambda_0 \approx 3$ kpc and the large-distance exponential decay length $\mu_0^{-1} \approx 17$ kpc, have already been determined from the study of the rotation curves of spiral galaxies as well as the internal dynamics of clusters of galaxies~\cite{NL6}. Therefore, it remains to determine the short-distance parameter $a_0$ and decide between $q_1$ and $q_2$. As a first step, preliminary lower limits can be placed on $a_0$  on the basis of current data regarding planetary orbits in the Solar System. For instance, for Saturn, a preliminary lower limit of  $a_0 \gtrsim 2 \times 10^{15}$  cm can be established for $q_1$, while   $a_0 \gtrsim 5.5 \times10^{14}$ cm for $q_2$. 

It has recently  been argued that the extension of the Tohline-Kuhn force~\eqref{III3} within the Solar System can likely be ruled out by  current observational data~\cite{Io, DX}. On the other hand, nonlocal gravity in the Solar System is characterized by the short-distance parameter $a_0$ and the associated nonlocal force  is in fact  different from the Tohline-Kuhn force. In Ref.~\cite{NL6}, which was primarily devoted to the study of the effective dark matter in galaxies and clusters of galaxies, the implications of $q_0$ for the Solar System were also considered for the sake of completeness; however, the short-range behavior of $q_0$ is the same as in the Tohline-Kuhn approach. Indeed, previous studies in this direction---see Refs.~\cite{FPC, NL3, NL6, LLY} and the references cited therein---have been confined to the Tohline-Kuhn force, which differs from the short-distance force of nonlocal gravity. In this connection, it is important to point out that nonlocal gravity is still in the early stages of development. To ameliorate this situation, the present paper has been devoted to a discussion of parameter $a_0$ of the reciprocal kernel and the short-distance behavior of nonlocal gravity in the Solar System.

\begin{acknowledgments}
BM is grateful to Lorenzo Iorio and Jeffrey Kuhn for valuable discussions.  
\end{acknowledgments}

\appendix{}
\section{Expansion of $\int_0^{\frac{\pi}{2}} \exp{(-x \sec \varphi)} \, d\varphi$ for $x\ge 0$ about $x=0$}

The integrals that we have encountered in our discussion of light deflection in nonlocal gravity all contain 
$\exp{(-x \sec \varphi)}$ for $x \ge 0$ in their integrands. Consider the simplest situation, namely, 
\begin{equation}\label{A1}
{\cal S}(x) = \int_0^{\frac{\pi}{2}} e^{-x \sec \varphi} \, d\varphi\,,
\end{equation}
which is a special case of Sievert's integral~\cite{A+S}. We note that ${\cal S}(0)=\pi/2$ and  ${\cal S}$ is not analytic about $x=0$. Moreover, 
\begin{equation}\label{A2}
{\cal S}'' -  {\cal S}= \int_0^{\frac{\pi}{2}} \tan^2 \varphi\, e^{-x \sec \varphi} \, d\varphi\,,
\end{equation}
where ${\cal S}'(x)=d{\cal S}/dx$, etc. The right-hand side of Eq.~\eqref{A2} can be evaluated using integration by parts; that is, 
\begin{equation}\label{A3}
 \int_0^{\frac{\pi}{2}} \sin \varphi\, e^{-x \sec \varphi} \, \frac{d}{d\varphi}\left(\frac{1}{\cos \varphi}\right)\, d\varphi =  -  {\cal S} + x\,({\cal S}'-{\cal S}''')\,.
\end{equation}
It follows that ${\cal S}'(x)$ satisfies the modified Bessel differential equation of order zero, namely,
\begin{equation}\label{A4}
 x\,H''(x) +H'(x)-x\,H(x)=0\,.
\end{equation}
The solutions of this equation are $I_0(x)$ and $K_0(x)$, which are the modified Bessel functions of order zero, see Ref.~\cite{A+S}. In fact,  $I_0(x)$, 
\begin{equation}\label{A5}
I_0(x)= \sum_{k=0}^{\infty} \frac{x^{2k}}{(2^k\, k!)^2}\,,
\end{equation}
is regular at $x=0$ and valid everywhere. Moreover, 
\begin{equation}\label{A6}
K_0(x)=-(\ln\frac{x}{2}+C)\, I_0(x) +  \sum_{k=1}^{\infty} \beta_k\, \frac{x^{2k}}{(2^k\, k!)^2}\,,
\end{equation}
where $C=0.577\dots$ is Euler's constant and
\begin{equation}\label{A7}
\beta_k= \sum_{n=1}^{k}  \frac{1}{n}\,.
\end{equation}

According to formula 27.4.3 on page 1000 of Ref.~\cite{A+S},
\begin{equation}\label{A8}
{\cal S}(x) = \int_{x}^{\infty} K_0(t)\, dt\,.
\end{equation}
It therefore follows that ${\cal S}' (x)= -K_0(x)$ and
\begin{equation}\label{A9}
{\cal S}(x) = \frac{\pi}{2} -  \int_{0}^{x} K_0(t)\, dt\,.
\end{equation}
Thus ${\cal S}(x)$ can be computed by substituting  Eq.~\eqref{A6} for $K_0$ in Eq.~\eqref{A9}.  To this end, it proves useful to define a new function $B(x)$, 
\begin{equation}\label{A10}
B(x):=-x +\int_0^x I_0(t)\,dt\,, 
\end{equation}
so that we have
\begin{equation}\label{A11}
B(x)= \sum_{k=1}^{\infty} \frac{x^{2k+1}}{(2k+1)\,(2^k\, k!)^2}\,.
\end{equation} 
It is then possible to use $I_0(x)=1+dB/dx$ in Eq.~\eqref{A6} and subsequently express Eq.~\eqref{A9} as 
\begin{equation}\label{A12}
{\cal S}(x)= \frac{\pi}{2} -x + x\,(\ln\frac{x}{2}+C) + \int_0^x (\ln\frac{t}{2}+C)\, \frac{dB}{dt}\, dt -  \sum_{k=1}^{\infty} \beta_k\, \frac{x^{2k+1}}{(2k+1)\,(2^k\, k!)^2}\,.
\end{equation}
Finally, we find via integration by parts that 
\begin{equation}\label{A13}
{\cal S}(x)= \frac{\pi}{2} -x + x\,(\ln\frac{x}{2}+C) +{\cal R}(x)\,,
\end{equation}
where
\begin{equation}\label{A14}
{\cal R}(x) = (\ln\frac{x}{2}+C)\, B(x) -  \sum_{k=1}^{\infty} \left(\beta_k+ \frac{1}{2k+1}\right)\, \frac{x^{2k+1}}{(2k+1)\,(2^k\, k!)^2}\,.
\end{equation}

\end{document}